\documentclass[letterpaper,english,reprint, prl]{revtex4-1}
\usepackage[latin9]{inputenc}
\usepackage{array}
\usepackage{booktabs}
\usepackage{multirow}
\usepackage{amsmath}
\usepackage{amssymb}
\usepackage{graphicx}
\usepackage{xcolor}

\makeatletter

\pdfpageheight\paperheight
\pdfpagewidth\paperwidth

\providecommand{\tabularnewline}{\\}

\makeatother

\usepackage{babel}
\begin{document}
\preprint{APS/123-QED}
\title{Exact Mapping Between a Laser Network Loss Rate and the Classical XY Hamiltonian by Laser Loss Control}  
\author{Igor Gershenzon}
\thanks{These authors contributed equally}
\affiliation{Department of Physics of Complex Systems, Weizmann Institute of Science, Rehovot 7610001, Israel}
\author{Geva Arwas}
\thanks{These authors contributed equally}
\affiliation{Department of Physics of Complex Systems, Weizmann Institute of Science, Rehovot 7610001, Israel}
\author{Sagie Gadasi}
\thanks{These authors contributed equally}
\affiliation{Department of Physics of Complex Systems, Weizmann Institute of Science, Rehovot 7610001, Israel}
\author{Chene Tradonsky, Asher Friesem, Oren Raz and Nir Davidson}
\email{nir.davidson@weizmann.ac.il}

\affiliation{Department of Physics of Complex Systems, Weizmann Institute of Science, Rehovot 7610001, Israel}

\begin{abstract}
Recently, there has been growing interest in the utilization of physical systems as heuristic optimizers for classical spin Hamiltonians. A prominent approach employs gain-dissipative optical oscillator networks for this purpose. Unfortunately, these systems inherently suffer from an inexact mapping between the oscillator network loss rate and the spin Hamiltonian due to additional degrees of freedom present in the system such as oscillation amplitude. In this work, we theoretically analyze and experimentally demonstrate a scheme for the alleviation of this difficulty. The scheme involves control over the laser oscillator amplitude through modification of individual laser oscillator loss. We demonstrate this approach in a laser network classical XY model simulator based on a digital degenerate cavity laser. We prove that for each XY model energy minimum there corresponds a unique set of laser loss values that leads to a network state with identical oscillation amplitudes and to phase values that coincide with the XY model minimum. We experimentally demonstrate an 8 fold improvement in the deviation from the minimal XY energy by employing our proposed solution scheme.
\end{abstract}

\maketitle

\section{\label{sec:introduction}Introduction}

Optimization problems are at the heart of numerous fields of science and industry from drug discovery \cite{Kuntz1992} to artificial intelligence \cite{Hopfield1982}. Unfortunately, many of the problems found in these applications are proven to be in the NP complexity class rendering their solution impractical even at modest input size \cite{garey1979computers}. Due to the significant applicability of these problems, various computational approaches have been developed to find practically useful approximations of their solutions in polynomial time \cite{vazirani2013approximation},\cite{talbi2009metaheuristics}. These approximation algorithms include non-linear programming \cite{luenberger1984linear}, semidefinite programming \cite{vandenberghe1996semidefinite}, local search algorithms \cite{aarts2003local}, evolution inspired algorithms \cite{zames1981genetic}, and physically inspired heuristic algorithms \cite{yang2014nature} among others.

Physically inspired algorithms are typically heuristic algorithms based on a mapping between the cost function and a physical energy landscape. Optimization is then carried out by mimicking the dynamics of physical systems towards low energy states \cite{aarts1988simulated}. In physical terms, the optimization problem is converted to a ground state search problem. For example, the simulated annealing (SA) algorithm mimics the cooling of metals by stochastic dynamics \cite{Kirkpatrick1983}, the quantum annealing algorithm mimics quantum dynamic of the ground state evolution \cite{das2005quantum}, the particle swarm algorithm mimics the manner in which avian flocks find food sources through distributed non-linear dynamics \cite{kennedy1995particle}. Extensive literature exists on the use and benchmarking of these algorithms for various applications \cite{Vesterstrom2004},\cite{Parejo2012}.

An alternative to ground-state search algorithms implemented on digital computers is the realization of specialized hardware setups. Prominent examples of systems for ground-state search include dedicated hardware for neural network training \cite{Steinkraus2005},\cite{Owens2008},\cite{Larger2012},  photonic Ising machines and XY simulators \cite{McMahon2016},\cite{Inagaki2016},\cite{Takeda2017},\cite{Nixon2013},\cite{Okawachi2019}, superconducting qubit annealing machines \cite{Boixo2014},\cite{das2005quantum}, polariton based XY simulators \cite{Berloff2017},\cite{Lagoudakis2017}, electronic Ising machines \cite{Yamaoka2015},\cite{Mahboob2016},\cite{Wang2019}, memristor network systems \cite{Thomas2013},\cite{adamatzky2013memristor},\cite{Stathis2014} and others. Many of these systems are aimed at finding the ground state of classical spin models. This is of special interest since many NP-complete problems can be mapped to such models \cite{Lucas2014}. Recent results on the universality of these models provide additional flexibility in mapping a given optimization problem to a given spin model \cite{DeLasCuevas2016}. 

Generally, two main ingredients are required of a physical ground state finder: (i) A correspondence between the physical system's stable states and the optimization function i.e. model energy landscape minima, (ii) A mechanism that ensures the evolution of the system to a stable state corresponding to an optimum of the optimization problem i.e the model ground state. 

An exact correspondence between a system's stable state and a model energy minimum would require the elimination of all physical degrees of freedom (DOF) absent from the model. For example, a continuous scalar DOF might be mapped to a discrete spin DOF. Achieving an exact correspondence is thus challenging both from the theoretical and experimental aspects \cite{Wang2013}, \cite{Leleu2019}, \cite{Kalinin2018}. However, additional DOFs can present an opportunity to improve the optimization success rate by embedding the model dynamics in the higher dimensional dynamics of the physical system. For example, such embedding could help to avoid trapping in local minima \cite{vandenberghe1996semidefinite},\cite{Luo2010}.

Ensuring the evolution of a physical system to its global ground state poses a significant challenge for complex non-convex energy landscapes \cite{mezard1987spin}. It is highly unlikely that physical systems can find the global ground state of such Hamiltonians in sub-expnonential time \cite{Aaronson2008},\cite{Aaronson2005}. Thus the main research question is whether physical optimization machines can outperform digital computers at these hard tasks by harnessing additional resources. Such resources might include short iteration times \cite{Tradonsky2019}, quantum tunneling and coherence \cite{das2005quantum}, inherent parallelism \cite{Tradonsky2019}, favorable scaling \cite{MacFaden2017}, or other resources \cite{Yamamoto2017}. 

In gain-dissipative optical oscillator networks one aims to find the ground state of classical spin models such as the Ising model and the XY model. In such systems, the phase of each optical oscillator (either OPO or laser) is mapped to a spin DOF. The idea put forward in \cite{Marandi2014}, is to utilize the mode-competition property of optical oscillators to select the network state with the lowest loss rate. In principle, there exists an external driving rate (pump) range for which only the mode associated with the lowest loss is a stable fixed point. Tuning the pump to this range could in principle result in finding the ground state of the spin model. Several types of optical oscillator networks based on this operation principle have been implemented and their efficacy at finding low energy states of various instances of spin models was studied \cite{Marandi2014},\cite{Inagaki2016},\cite{McMahon2016}. 

It was shown that an approximate mapping can be established between the spin model energy and the oscillator network loss rate when the inter-oscillator coupling is low \cite{Wang2013}. In this regime, the oscillation amplitudes are almost the same for all oscillators. This reduces the loss rate of the network to an equivalent spin model energy \cite{Wang2013}. On the other hand, low coupling strength leads to small energy gaps and thus long relaxation times to the ground state. Additionally, this leads to increased sensitivity to imperfections and noise, dictating the use of finite-size coupling in practical applications. This, in turn, leads to unequal (heterogeneous) oscillation amplitudes which preclude the exact mapping between the loss rate and the spin model energy. Since this limitation stems from a finite size coupling between oscillators, it is inherent to any coupled oscillator network. Several theoretical approaches for mediating this effect have been suggested \cite{Leleu2019},\cite{Kalinin2018}. Both methods rely on equalizing the amplitude of the oscillators by imposing additional dynamical equations on each oscillator. To the author's knowledge, such techniques haven't been experimentally demonstrated and studied to date. 

In this work, we theoretically analyze and experimentally demonstrate the problem of unequal amplitudes.  This is carried out on a simple laser network acting as an XY model ground state finder. We find that an inherent contradiction exists between finding the ground state in the vicinity of the oscillation threshold and accurately mapping the network state to an XY state. To solve this, we devise and experimentally demonstrate a scheme for the solution of the amplitude heterogeneity problem in a laser network. The implemented solution is based on controlling the individual oscillator loss rates to achieve equal oscillation amplitudes. The phase in such states is found to correspond to XY model minima. We prove that for each XY model minimum, the set of laser network parameters for which the laser network phase values coincide with the XY model is unique. We achieve this without augmentation of the intrinsic oscillator dynamics.

\section{\label{sec:problem_presentation}Problem Presentation}

We illustrate the amplitude heterogeneity problem and our solution to it with a simple laser network - the "house graph" \cite{west2001introduction} with negative (anti-ferromagnetic) couplings, shown in figure \ref{fig:House model}(a). The classical XY Hamiltonian for the house graph is given by \begin{equation} 
H_{\mathrm{XY}}=-\sum_{nm}\kappa_{nm}\cos\left(\phi_{n}-\phi_{m}\right)\label{eq:XY energy} 
\end{equation} where $\kappa_{nm}$ is the coupling matrix defined by the house graph and $\phi_{n}$ is the phase of the $n$th spin (or oscillator). The ground state of $H_{\mathrm{XY}}$ for this graph can be found analytically (see supplementary materials) and is plotted in figure \ref{fig:House model}(a). The minimal loss state of a network of identical laser oscillators \cite{Nixon2013} is calculated (see section \ref{sec:Theoretical_analysis}) and plotted in figure \ref{fig:House model}(b). As seen, The amplitude of the lasers is highly heterogeneous where the uppermost laser completely shuts down in spite of having identical gain, loss and frequency to the other four lasers. This is due to the frustration in the triangular facet of the house graph \cite{Diep2013}. The resulting phases of the minimal loss state deviate significantly by as much as $43.2^{\circ}$ from the XY ground state (in the phase difference denoted by $\Delta\phi$ in figure \ref{fig:House model}(a)). This example highlights the effect of the inexact mapping between the XY model and the laser network loss rate due to additional DOFs: the oscillation amplitudes. 

We propose to overcome this effect by tuning the laser oscillator parameters. As shown in section \ref{sec:Theoretical_analysis}, if the loss of each laser is adjusted such that the  minimal loss state has uniform intensities for all lasers, the phases corresponding to the XY ground state are exactly recovered. This verifies the exact mapping between the minimal loss state of uniform amplitude lasers to the XY ground state.

\section{\label{sec:experimental_results}Experiment}

\begin{figure}

\includegraphics[scale=0.51]{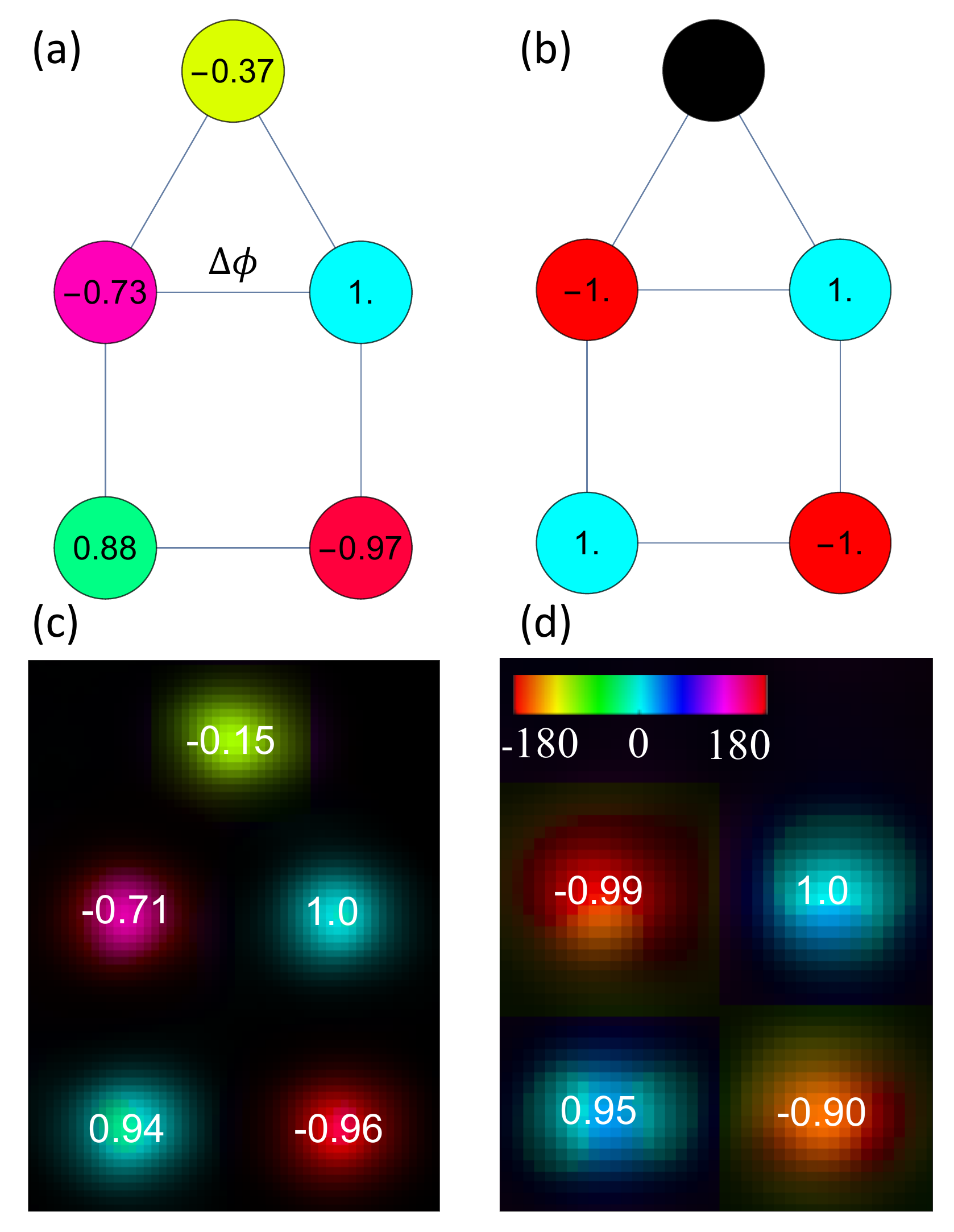}

\caption{\label{fig:House model}(a) The calculated anti-ferromagnetic XY model ground state on the house graph.  (b) The calculated laser network minimal loss state on the house graph (slightly above the oscillation threshold).  (c) A measured laser network state with adjusted loss values to achieve uniform oscillation amplitudes is in good agreement with the calculated state shown in (a). The (weak) coupling strength is $\kappa \approx 0.05\alpha_{0}$ and the pump is $P-P_\mathrm{th} \approx \kappa$. (d) A measured minimal loss state of a network of identical laser oscillators is in good agreement with the calculated state shown in (b). The (strong) coupling strength is $\kappa \approx 0.4\alpha_{0}$ and the pump value is slightly above the network oscillation threshold ($P-P_\mathrm{th} \approx \kappa/4$). Color hue depicts phases according to color bar in (d) and color brightness depicts the laser amplitude where black corresponding to zero. The phase cosine is written explicitly for each laser. }
\end{figure} 

\begin{figure}
\includegraphics[scale=0.3]{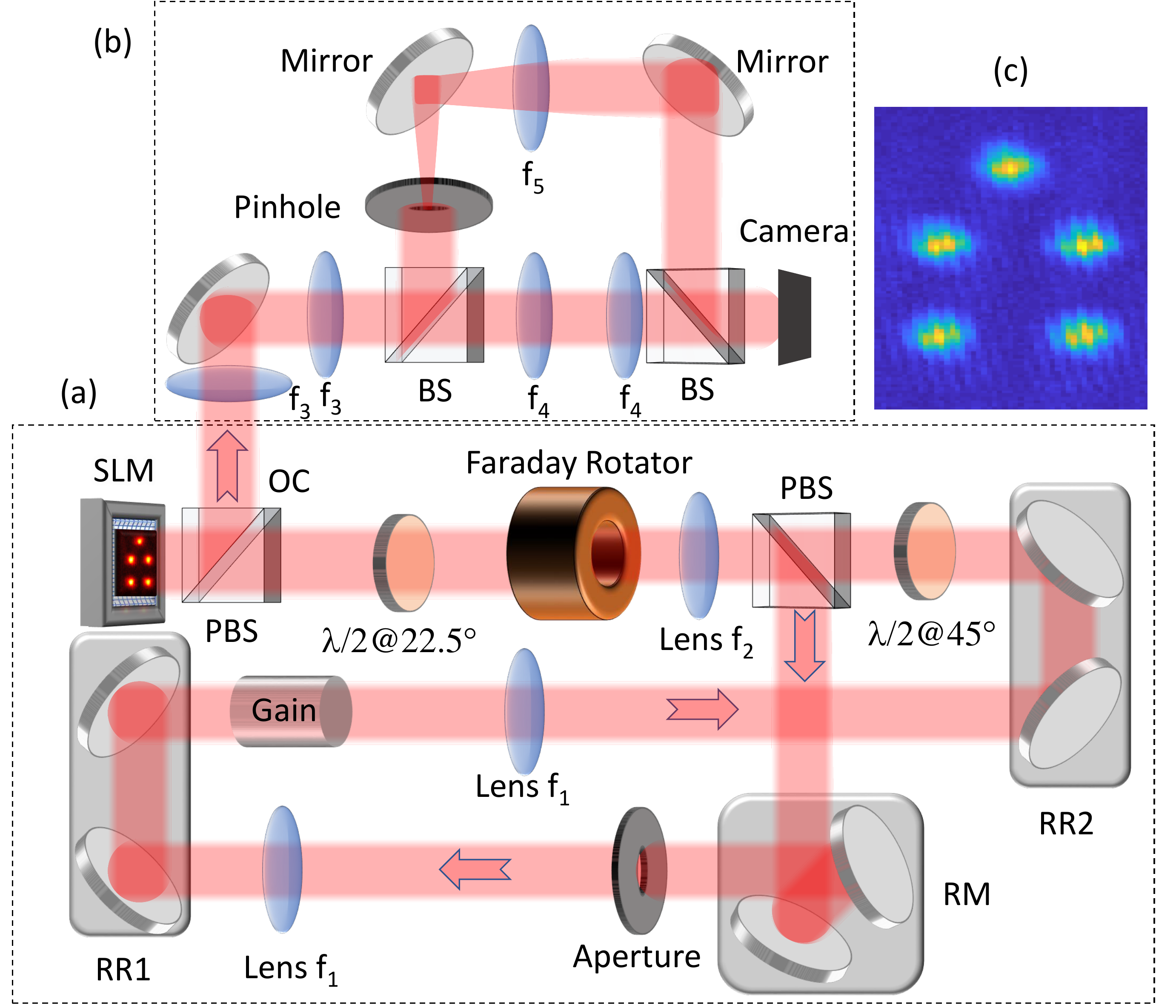}

\caption{\label{fig:exp. setup}Schematic illustration of the experimental setup used to support a laser oscillator network and to measure its state. (a) folded ring degenerate cavity laser supporting the oscillator network (b) Interferometer for laser network intensity and phase measurement. (c) The detected interference fringes for a house laser network at uniform amplitude. Abbreviations: SLM - spatial light modulator, OC - output coupler, PBS - polarizing beam splitter, RR1 and RR2 - retro reflecting mirrors, RR2 - right retro reflecting mirrors, $\lambda/2$ - half wave plate, BS - beam splitter, RM - reflector mirrors.}
\end{figure}

The laser network used to simulate the XY model is implemented in a digital degenerate ring cavity laser  \cite{Tradonsky2019},  schematically depicted in figure \ref{fig:exp. setup}. The cavity includes two 4f imaging telescopes, an Nd:YAG gain medium, a spatial light modulator (SLM), an optical isolator and an output coupler. The gain medium is pumped by a Xenon flash lamp, resulting in $200\mu s$ quasi-CW pulses. The pumping rate is controllable by changing the voltage of the flash-lamp activation pulse. The 8f telescopes image the field distribution from the SLM onto itself after a cavity roundtrip \cite{Arnaud1969}. For more details, see supplementary information. 

The SLM is utilized as a digital mirror whose complex-valued reflectivity at each pixel is controlled \cite{Ngcobo2013}. This control is used to generate a network of single mode lasers with any desired geometry ("house" network here) by imposing  lasing only on specific spots where a non-zero controllable reflectivity is defined. This adjustable reflectively is then used to control the loss of each laser independently via a closed loop control scheme. We first use this scheme to compensate for loss and gain imperfections in our system and to form a network of identical laser oscillators and later we use it to form a network of lasers with uniform amplitude (see methods section). We adjust the phase of the reflectively for each laser in the network independently with an additional close loop control scheme to compensate for aberrations in our laser cavity and ensure identical frequency for all lasers as will be described in a future publication.

In the degenerate cavity used in this work, each laser spot corresponds to an independent laser oscillator \cite{Nixon2011}. Two methods are applied in this work to introduce diffractive coupling between lasers, (i) a circular aperture placed at the Fourier plane of the second 4f telescope as depicted in figure \ref{fig:exp. setup}(a) generates weak coupling \cite{Tradonsky2017} and (ii) a lens placed instead of the aperture, generates strong (Talbot) coupling \cite{Mahler2019}, See supplementary material for more details. 

The measurement of the laser network phase and amplitude state is carried out by using an interferometric apparatus schematically depicted in figure \ref{fig:exp. setup}(b). On one arm of the interferometer, a pinhole and a lens serve to select and expand a single laser. In the second arm, a 4f telescope is used to directly image the laser field distribution on the SLM. The light from both arms is recombined on a CCD detector resulting in interference fringes on each laser spot (see figure \ref{fig:exp. setup}(c)). 

Figure \ref{fig:House model}(d) depicts the measured state of a network of identical lasers with anti-ferromagnetic coupling. The coupling strength (where a bond exists) is $\kappa \approx 0.4\alpha_{0}$ and the pump value, $P$ is slightly above the network oscillation threshold ($P-P_\mathrm{th} \approx \kappa/4$) where $\alpha_{0}$ is the single oscillator loss rate and $P_\mathrm{th}$ is the threshold pump value. In an extreme manifestation of amplitude heterogeneity, the uppermost laser is "off" due to its frustrating coupling, despite having identical loss to the other lasers. The four lower lasers assume a simple anti-ferromagnetic ring configuration which does not correspond to the XY ground state. Quantatively, a deviation of $35^{\circ}$ is observed in the value of $\Delta\phi$ (corresponding to $0.26$ deviation in $\cos{\left(\Delta\phi\right)}$. The measured network state is in good agreement with the theoretical prediction in figure \ref{fig:House model}(b).

Next, we adjust the loss of each laser such that all lasers have the same amplitude (while maintaining identical frequencies and pump values). The loss rate modification of laser to was found to be 0.204 for the uppermost laser (in units of inverse cavity round-trip time) and -0.041,0.032,0.062,0.0014 for the lower four (starting from the upper left laser and going counter-clockwise) and we denote it as the modified loss vector ${\vec{\Delta\alpha} = \left(0.204,-0.041,0.032,0.062,0.0014\right)}$. The loss modification is directly determined from the SLM reflectivity value \cite{Siegmann1986}.

Figure \ref{fig:House model}(c) depicts the state of a uniform amplitude laser network with weak anti-ferromagnetic coupling ($\kappa \approx 0.05\alpha_{0}$) and a pump value of $P - P_\mathrm{th} \approx \kappa$. The laser amplitudes are seen to be uniform while the phase values indeed correspond to the ground state of the XY model. Quantitatively, the deviation in $\Delta\phi$ is $2^{\circ}$. 

\begin{figure}
\includegraphics[scale=0.52]{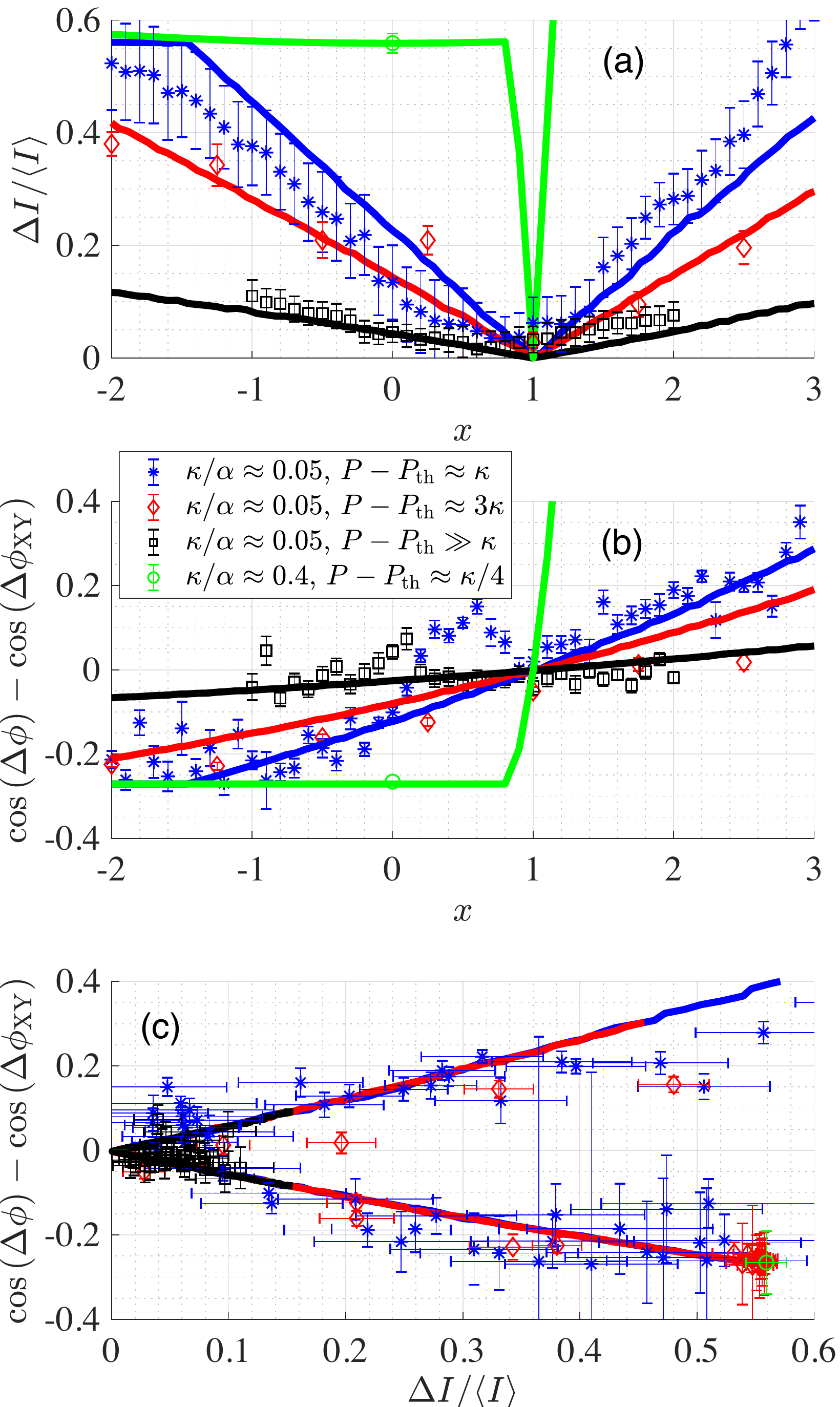}

\caption{\label{fig:network-state-vs-delta-alpha}House network state as a function of the laser loss at various operation regimes - comparison of measurements (markers) to theory (solid lines) for all cases of coupling strength and pump strength of Table I. Error bars are estimated as $67\%$ confidence intervals. (a) The amplitude heterogeneity versus $x$.  (b) The phase deviation from the XY ground state phase versus $x$. (c) The phase deviation from the XY ground state phase versus the amplitude heterogeneity for all cases.}

\end{figure}

Next, we investigate the network state as a function of the amplitude heterogeneity, we scan the laser loss vector by interpolating between the identical oscillator and the uniform amplitude states in the following manner 

\begin{equation}
\vec{\alpha}\left(x\right)=\vec{\alpha}_{0}+x\vec{\Delta\alpha} \label{eq: loss space trajectory}
\end{equation} where $x$ is the interpolation parameter in loss space and $\vec{\alpha}_{0}$ is a vector of identical single oscillator loss values $\alpha_{0}$. The identical loss and uniform amplitude points are reached at $x=0$ and $x=1$, respectively. The laser network state is measured as the network is driven along this trajectory for different combinations of pump and coupling strength values as summarized in table \ref{tab:result-summary}. The experimental results and the corresponding theoretical results are plotted versus $x$ in figure \ref{fig:network-state-vs-delta-alpha}. The network state is quantified by the normalized intensity heterogeneity, $\Delta I/\langle I\rangle$ (see figure \ref{fig:House model}(a)) where $I$ is the squared amplitude and by the deviation of the phase difference cosine $\cos{\left(\Delta\phi\right)}$  from its value for the XY ground state $\cos{\left(\Delta\phi_{\mathrm{XY}}\right)}$ (see figure \ref{fig:House model}(b)). The theoretical lines were obtained by simulating the coupled laser rate equations \cite{Rogister2004}.  The pump value, for each measurement set, is evaluated by finding the best fit for the experimental results.

\begin{table}
\begin{tabular}{|c|c|c|c|c|c|c|c|}
\hline
\toprule 
\multirow{2}{*}{} & \multirow{2}{*}{$\frac{\kappa}{\alpha}$} & \multirow{2}{*}{$P-P_{\mathrm{th}}$} & \multirow{2}{*}{} & \multicolumn{2}{|c|}{$x=0$} & \multicolumn{2}{|c|}{$x=1$}\tabularnewline
\cmidrule{5-8} \cmidrule{6-8} \cmidrule{7-8} \cmidrule{8-8} 
 &  &  &  & $\Delta\phi[^{0}]$ & $\Delta I/I[\%]$ & $\Delta\phi[^{0}]$ & $\Delta I/I[\%]$\tabularnewline
\midrule 
\multirow{2}{*}{i} & \multirow{2}{*}{$5\%$} & \multirow{2}{*}{$\kappa$} & E & $146\pm1$ & $13.4\pm0.4$ & $135\pm1$ & $6.2\pm0.1$\tabularnewline
\cmidrule{4-8} \cmidrule{5-8} \cmidrule{6-8} \cmidrule{7-8} \cmidrule{8-8} 
 &  &  & T & $148.3$ & $22.4$ & $136.8$ & $0$\tabularnewline
\midrule 
\multirow{2}{*}{ii} & \multirow{2}{*}{$5\%$} & \multirow{2}{*}{$3\kappa$} & E & $150\pm1$ & $20.9\pm0.3$ & $141\pm1$ & $2.7\pm0.1$\tabularnewline
\cmidrule{4-8} \cmidrule{5-8} \cmidrule{6-8} \cmidrule{7-8} \cmidrule{8-8} 
 &  &  & T & $143.9$ & $14.2$ & $136.8$ & $0$\tabularnewline
\midrule 
\multirow{2}{*}{iii} & \multirow{2}{*}{$5\%$} & \multirow{2}{*}{$\gg \kappa$} & E & $133\pm1$ & $4\pm0.04$ & $140\pm2$ & $3\pm0.04$\tabularnewline
\cmidrule{4-8} \cmidrule{5-8} \cmidrule{6-8} \cmidrule{7-8} \cmidrule{8-8} 
 &  &  & T & $139.0$ & $4.2$ & $136.8$ & $0$\tabularnewline
\midrule 
\multirow{2}{*}{iv} & \multirow{2}{*}{$40\%$} & \multirow{2}{*}{$\kappa/4$} & E & $174\pm4$ & $55.9\pm 2$ & - & -\tabularnewline
\cmidrule{4-8} \cmidrule{5-8} \cmidrule{6-8} \cmidrule{7-8} \cmidrule{8-8} 
 &  &  & T & $180$ & $56$ & $136.8$ & $0$\tabularnewline
\bottomrule
\hline
\end{tabular}

\caption{\label{tab:result-summary}Experimental and theoretical result summary for all measured cases for identical oscillator network ($x=0$) and uniform amplitude ($x=1$) states. E and T denote experimental and theoretical results, respectively}
\end{table}

The experimental and theoretical values for the phase and amplitude heterogeneity for all measured cases are summarized in table \ref{tab:result-summary} at the identical oscillator and uniform amplitude states. Qualitatively it is seen in figure \ref{fig:network-state-vs-delta-alpha}(b) that the deviation from the XY model ground state phase changes from negative values (corresponding to a phase $\Delta\phi=180^{\circ}$) at negative $x$, where the uppermost laser is off, to zero at $x=1$ where the amplitude is uniform. The deviation rises from zero for $x>1$ and uppermost laser amplitude is higher than the lower four. Good agreement is found between the experimental results and the stimulative prediction for all measurement cases. Figure  \ref{fig:network-state-vs-delta-alpha}(a) reveals that, as designed, the amplitude heterogeneity is minimized at $x=1$ for all measurement sets. Figure \ref{fig:network-state-vs-delta-alpha}(b) shows that, indeed, at the uniform amplitude point the laser phase is in good agreement with the XY model phase. It is also evident that the phase value deviates from the XY minimum phase as the amplitude heterogeneity increases. The deviation from uniform amplitude is seen to be steeper as the pump value is closer to the oscillation threshold. Accordingly, the laser network phase value deviates from the XY phase more significantly for pump values closer to threshold. On the other hand, at gain values significantly larger than the oscillation threshold the amplitude heterogeneity is small and the phase is almost constant at the XY model value. 

Figure \ref{fig:network-state-vs-delta-alpha}(c) directly plots the deviation from the XY ground state phase versus the amplitude heterogeneity. It is seen that both theoretically and experimentally, the data collapses to a single curve which indicates that in this example, the phase deviation depends only on the amplitude heterogeneity. Focusing on the unmodified laser network fixed point ($x=0$), it is seen that higher coupling strength leads to significantly larger amplitude heterogeneity and consequently extreme phase deviations from the XY model phase exceeding $40^{\circ}$. Unfortunately, the correction of this extreme heterogeneity by amplitude feedback was experimentally unavailable due to high loss values present at strong coupling strength \cite{Tradonsky2017}. The theoretical plot for strong coupling values and pump close to threshold reveals a very sharp transition from extreme amplitude heterogeneity to uniform intensity as a function of $x$. 

\begin{figure}
\includegraphics[scale=0.5]{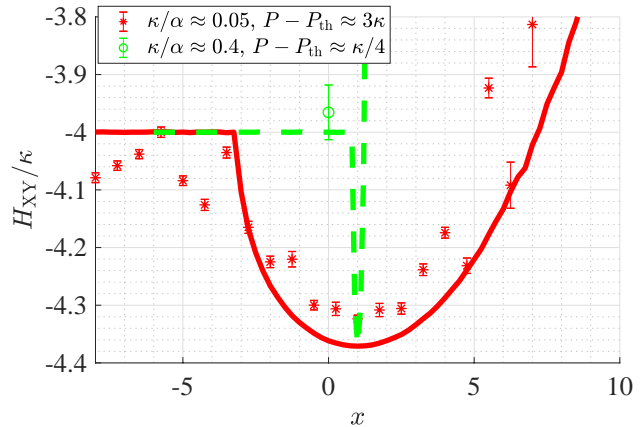}

\caption{\label{fig:XY-model-energy}Calculated XY model energy estimation from laser network fixed point phase versus $x$ - comparison of measurement (markers) to theory (solid lines). Error bars are estimated as $67\%$ confidence intervals.}
\end{figure}

The final test of the effect of amplitude heterogeneity on the performance of a laser network as an XY model optimizer is the effect on the XY energy (eq. \ref{eq:XY energy}) calculated from the measured laser network phase values. The energy calculated from measured phase values and from simulation results is plotted in figure \ref{fig:XY-model-energy} versus $x$ for two of the cases in table \ref{tab:result-summary}.  Figure \ref{fig:XY-model-energy} reveals that the theoretical minimum of the XY model energy is obtained to the best approximation at the uniform amplitude point $x=1$. A good general agreement is found between the measured values and the simulation across the whole $x$ range. In addition, it is evident again that at high coupling strength the deviation from the XY energy minimum is more severe at the identical oscillator network state and that generally the dependence on $x$ is steeper around the uniform intensity point. 

\section{\label{sec:Theoretical_analysis}Analysis}

In the following we show that the proposed method of oscillator loss tuning indeed gives rise to uniform amplitude solutions of the laser rate equations. Moreover, we show that the loss values for each XY minimum are unique. The analysis starts from the coupled laser rate equations for $M$ class-B, identical frequency, laser oscillators \cite{Rogister2004}
\begin{eqnarray}
 \frac{d E_m}{dt} & =&\frac{1}{\tau_p} \left[ (G_m-\alpha_m) E_m - \sum_{n} \kappa_{mn} E_n \right]\label{eq: field dynamics}\\
  \frac{dG_{m}}{dt}& =& \frac{1}{\tau_{c}}\left[P_{m}-G_{m}\left(1+\frac{\left|E_{m}\right|^{2}}{I_{\mathrm{sat}}}\right)\right]\label{eq: gain dynamics}
\end{eqnarray}
where $E_{m}(t)$ is the slowly-varying electric field of the $m$th laser, $G_{m}(t)$ is the $m$th active medium gain, $\tau_{p},\tau_{c}$ are the cavity round trip time and gain medium fluorescence lifetime respectively, $I_{\mathrm{sat}}$ is the gain medium saturation intensity, $\alpha_{m}$ and $P_{m}$ are the normalized loss coefficient and active medium pump rate of the $m$th laser. The coupling matrix element $\kappa_{mn}$ is the normalized field injection coefficient from the $n$th laser to the $m$th laser. For the theoretical analysis it is convenient to write the equations in a dimensionless form. This is done by rescaling the units of the electric field and time by setting $I_{\mathrm{sat}}=1$ and $\tau_{p}=1$. 

First, we establish the connection between the laser network and the classical XY model. To this end we assume that the oscillation amplitude is an identical constant i.e. $A_{m}(t)=\left|E_{m}(t)\right|=A,\forall m$. By setting  $E_{m}(t) = A e^{i\phi_{m}}$ in eq. \ref{eq: field dynamics} and looking for steady state solutions we get \cite{Tamate2016}, \cite{Berloff2017}
\begin{equation}
    \partial H_{\mathrm{XY}}/\partial\phi_{m}=0 
\end{equation}
where $H_{\mathrm{XY}}$ is the classical XY model Hamiltonian defined in eq. \ref{eq:XY energy}. Thus, the phase values at uniform intensity fixed points correspond to the phase values of the XY model extremal points i.e. an exact correspondence is reached between the laser network loss and the XY energy.

To achieve this exact correspondence, the inherent amplitude heterogeneity has to be addressed. In the following, we outline our proposed scheme for achieving this by tailing the loss of each oscillator  $\alpha_{m}$. Using eq. \ref{eq: field dynamics}, \ref{eq: gain dynamics} and setting $E_{m}(t) = A_{m}e^{i\phi_{m}}$, the steady state condition reads
\begin{equation}
\sum_{n}\kappa_{mn}A_{n}\cos(\phi_{n}-\phi_{m})=\left(\frac{P_{m}}{1+A_{m}^{2}}-\alpha_{m}\right)A_{m}.\label{eq: laser  network fixed point}
\end{equation}
Now suppose that we have modified the loss, such that all lasers now have the same amplitude $A_{m}=A,\forall m$. It immediately
follows that:
\begin{align}
\Delta\alpha_{m} & =-\sum_{n}\kappa_{mn}\cos(\phi_{n}-\phi_{m})+\delta\label{eq: loss phase relation}\\
 & A=\sqrt{\frac{P}{\alpha+\delta}-1}
\end{align}
denoting the loss as $\alpha_{m}=\alpha_{0}+\Delta\alpha_{m}$, where $\Delta\alpha_{m}$ is the loss modification and where we have assumed that $P_{m}=P,\forall m$. This set of equations directly relates the XY solutions to the added loss. For each XY fixed point with phase values $\phi_{m}$ there is a unique (up to the constant $\delta$ that will only change $A$) loss pattern $\Delta\alpha_{m}$ for which the amplitudes are uniform. 

Let us now demonstrate our approach for the house graph example. Using eq. \ref{eq: loss phase relation}, the added loss which corresponds to the XY ground state is uniquely defined by $\Delta\alpha_{m}\approx\kappa\left(1.2,-0.13,0,0, -0.13\right)$, see the supplementary materials for additional details. 

To directly show the correspondence, note that summing up eq. \ref{eq: loss phase relation}, we obtain:
\begin{equation}
H_{XY}=\sum_{m}\Delta\alpha_{m}-M\delta
\end{equation}
Thus, the added loss is directly related to the XY model energy at the fixed point. Moreover, the XY energy of the state can be inferred solely from the loss modification values $\Delta\alpha_{m}$. 

To understand the extreme intensity heterogeneity found near the oscillation threshold, we observe eq. \ref{eq: field dynamics} slightly above the oscillation threshold. Since the lasing transition is a super-critical pitchfork bifurcation \cite{mandel2005theoretical}, we can assume $\left|E_{m}\right|\ll1\ \forall m$ and arrive at the following eigenvalue equation for the fixed points:
\begin{equation}
\left(\hat{\alpha}+\hat{\kappa}\right)\mathbf{E}\ =\ \hat{P}\mathbf{E}
\end{equation}
where $\hat{\alpha}$ and $\hat{P}$ are the diagonal matrices of the $\alpha_{m}$ and $P_{m}$ vectors, $\hat{\kappa}$ is the coupling matrix and $\mathbf{E}$ is a vector of complex valued electric fields. Thus, the lasing threshold $P_{\mathrm{th}}$ is given by the lowest eigenvalue of $\left(\hat{\alpha}+\hat{\kappa}\right)$, and the lasing mode at the threshold is the corresponding eigenvector $\mathbf{E}_{\mathrm{min}}$. Generally, unless a special symmetry is present in the problem, the eigenvector has an arbitrary intensity heterogeneity. In the house graph example, this eigenvector is given by $\mathbf{E}_{\mathrm{min}}^{\mathrm{House}}=\left(0,1,-1,1,-1\right)$ as depicted in figure \ref{fig:House model}(b),(d) i.e. the roof oscillator does not turn on which is an extreme example of this phenomenon. Moreover, it can be easily shown that slightly above the network oscillation threshold, for generic real symmetric coupling matrices, the lasing phase is invariably either zero or $180^{\circ}$. This is explained by the algebraic fact that the eigenvectors of real symmetric matrices are real valued \cite{horn2012matrix}.

\section{\label{sec:conclusions}Conclusions}
We have demonstrated and studied the effects of amplitude heterogeneity on the performance of a laser oscillator network XY simulator. The effect was studied on a small and simple laser network - the house graph with anti-ferromagnetic coupling. It was found that amplitude heterogeneity is the most severe at the minimum loss state (infinitesimally above oscillation threshold) and that the phase values at this state deviate the most form the XY model ground state phases. We proposed and demonstrated a scheme for the solution of the amplitude heterogeneity effect. The scheme involves changing the laser oscillator loss rate to achieve uniform amplitudes. We found that upon equalization of the amplitudes the laser phases recover the XY ground state phases. We also proved that the set of loss values equalizing the amplitudes is unique for each XY model minimum.

\section{\label{sec:Methods}Methods}

\paragraph{\label{subsec:phase_measurement}Phase measurement}
Since multiple lasing frequencies (multiple longitudinal modes) coexist in our system, the measured interference fringes (figure \ref{fig:exp. setup} (b),(c)) amplitude and phase i.e. the complex valued coherence \cite{goodman2015statistical} corresponds to an ensemble average over the coherence \cite{Pal2019}. To interpret these ensemble averaged results, note that for time-reversal symmetric systems (identical frequency oscillators with symmetric real valued coupling matrix), complex conjugation of each field solution also yields a valid solution. Hence, a state with phases $\phi_{m}$ would give rise to a measured coherence of $\left\langle C_{m}\right\rangle = \frac{1}{2}\left(e^{i\phi_{m}}+e^{-i\phi_{m}}\right)=\cos\left(\phi_{m}\right)$. Thus our interferometric setup directly measures the cosine of the phase as the amplitude of the interference fringes and the phase is obtained by $\phi_m = \arccos{\left(\left\langle C_{m}\right\rangle\right)}$.

\paragraph{\label{subsec:amplitude_control}Amplitude control}
Control over the laser intensities, to suppress intensity heterogeneity, is carried out by a laser loss closed loop control scheme as follows. Operating at quasi-CW mode, the pulse averaged intensity of each laser is measured using the imaging apparatus. This intensity is averaged over a pre-defined number of pulses after which the SLM reflectivity of each laser is modified according to a proportional control feedback \cite{bequette2003process} scheme with a constant target intensity for all lasers. This protocol is carried out until all intensities converge to the target value (up to a to a predefined tolerance). 

\paragraph{\label{subsec:frequency_control}Frequency control}
To achieve the required phase measurement resolution, the laser frequency heterogeneity i.e. detuning imperfection was also addressed. This imperfection is caused by phase aberrations and cavity misalignment which lead to systematic phase deviations in the laser network fixed points. To compensate for this imperfection, the oscillator phase is measured per pulse and controlled via the SLM reflectivity phase as will be described in a future publication.

\paragraph{\label{subsec:imperfection_compensation}Experimental imperfection compensation}
Due to experimental imperfections, making the oscillators identical entails the nullification of parasitic loss variation. To that end, we applied the intensity feedback scheme separately to two sub-graphs of the house graph: a square graph and a triangle graph. Since both subgraphs have uniform intensity fixed points when the oscillators are identical, finding the loss values that result in uniform intensity compensates for parasitic loss heterogeneity. The loss values resulting in uniform intensity are found by using the closed loop intensity feedback as described the previous paragraph.

\section{Acknowledgements}
O.R. is the incumbent of the Shlomo and Michla Tomarin career development chair, and is supported by the Abramson Family Center for Young Scientists and by the Israel Science Foundation, Grant No. 950/19 and 1881/17.

\bibliographystyle{IEEEtran} %
\bibliography{house_paper_bib}

\onecolumngrid
\appendix

\newpage
\section*{Supplementary Material}

\subsection{Materials - Detailed Experimental Setup Description}

The detailed experimental arrangement of the digital degenerate ring cavity laser (DDRCL)\cite{Tradonsky2019} is schematically presented in figure 1 (a) in the manuscript. The DDRCL includes a gain medium, two 4f telescopes with one common lens, a reflective phase only spatial light modulator (SLM), a coupling element in the far-field (FF): an aperture or a lens, two retroreflectors and pentaprism-like 90\textdegree{} reflector (all from high reflectivity mirrors), two polarizing beam splitters (PBS), two half-wave plates ($\lambda$/2) and a Faraday rotator. 

The laser gain medium was a 1.1\% doped Nd-YAG rod of 10-mm diameter and 11-cm length. For quasi-CW operation, the gain medium was pumped above threshold by a $200\mu s$ pulsed xenon flash lamp operating at 1700-1950v and a repetition rate of 1 Hz to avoid thermal lensing. Each 4f telescope consists of two plano-convex lenses, with diameters of $50.8 mm$ and focal lengths of $f1 = 750 mm$ and $f2 = 500 mm$ at the lasing wavelength of $1064 nm$. The SLM was Meadowlark (liquid crystal on silicon (LCOS)) with a zero order diffraction efficiency of 88 \%, an area of $17.6 mm$ by $10.7 mm$, 1920 by 1152 resolution, $9.2\mu m$ pixel size, and a high damage threshold.

In the DDRCL, each of the two 4f telescopes has one lens f1 and a common lens f2. The first telescope images the field distribution at the center of the gain medium onto the SLM where the reflectivity of each effective pixel \cite{Ngcobo2013} is controlled. The second telescope, which contains a coupling element in the FF, images the field distribution at the SLM back onto the gain medium. Since the SLM operates on axis and by reflection on horizontal polarized light, half of the ring degenerate cavity was designed as a twisted-mode \cite{Evtuhov1965} linear degenerate cavity \cite{Arnaud1969} and the other half as regular ring cavity laser \cite{Arnaud1969}. The two halves are connected by PBS1, which separates the two counter-propagating beams into two different cross-polarized paths. A large aperture Faraday rotator together with a half-wave plate (HWP) at 22.5\textdegree{} and another PBS2 (which also serves as an output coupler) enforce unidirectional operation of the DDRCL. A 90\textdegree{} reflector flips left and right areas of the beam. A second HWP at 45\textdegree{} rotates the polarization from vertical to horizontal to pass through PBS1. 

The detection arrangement is shown in figure 1 (b) in the manuscript, and includes a CMOS camera, lenses, beam splitters and a pinhole to form an interferometer. The first interferometer channel images the near field onto the camera with an 8-f telescope. In the second interferometer channel, one of the lasers is selected by the pinhole and expanded by an additional lens to serve as a reference beam. The two channels are then combined by a beam splitter, to interfere on the camera with a small relative angle.

The local reflectivity magnitude of the SLM is determined by local phase differences between adjacent pixels and affects the amount of light diffracted out of the cavity. The local reflectivity phase is determined by the local average phase of the adjacent pixels \cite{Ngcobo2013}. For example, adjacent pixels with phases of {[}0, 0{]} will result in high reflectivity and 0 phase, whereas adjacent pixels with phases of {[}0, $\pi${]} will result in no reflectivity and $\pi$/2 phase. The reflectivity pattern can form arbitrary loss and phase distribution, and it is used to create the lasers, and to compensate for the cavity's inhomogeneities and aberrations.

\subsection{Laser Coupling Matrix Design} 
The coupling between the lasers was achieved by placing a diffractive element in the FF plane. The diffractive elements alter the perfect imaging condition of the 8-f cavity such that a portion of the light from each laser leaks to the neighboring lasers \cite{Tradonsky2017}. 

In the experiment, for weak coupling ($\sim$0.05\%) a FF aperture was used, and for strong coupling ($\sim$0.4\%) a FF lens was used. The strength and sign of the coupling coefficients between neighboring lasers depend on the distance between the lasers, their diameter, and on the aperture diameter in case of FF aperture, or on the focal length in the case of FF lens. To determine these parameters such that the coupling between nearest neighbors (NN) will be negative, and the coupling to the other neighbors will be negligible, we turned to numerical simulations. The coupling coefficients were calculated by simulating the diffraction of a Gaussian mode after one round trip (RT) in the cavity \cite{Tradonsky2017}. Then the overlap integral of the propagated field with a transversely shifted Gaussian mode was calculated, and was normalized by the overlap integral of the propagated field with the original Gaussian mode.

\subsubsection{FF aperture}

\begin{figure}
\includegraphics[scale=0.5]{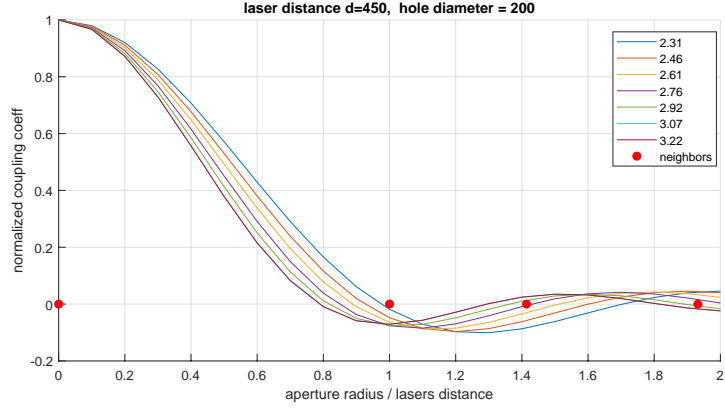}
\caption{\label{fig:ff aperture coupling coeff}The figure displays the coupling coefficient as a function of the normalized distance between the lasers, for different FF aperture diameters. The lasers distance is normalized by the NN distance.  The legend displays the diameters of the FF aperture in millimeter. The lasers diameter is $200\mu m$ and the NN distance is $450\mu m$. The bold red points mark the distances to the neighbors on the house graph.} 
\end{figure}

In the experiments where a FF aperture was used as the diffractive element, we kept the diameter of the lasers ($200\mu m$) and their distances ($450\mu m$) constant, and used numerical simulations to find the optimal aperture diameter. The optimal aperture diameter will result in maximal NN coupling strength (negative coupling) and negligible coupling to the other neighbors. Figure \ref{fig:ff aperture coupling coeff} displays the coupling coefficient as a function of the normalized laser distance, for different aperture diameters. The bold red points mark distances that correspond to distances between lasers on the house graph. It easy to see that the optimal aperture diameter is around $2.9mm$. 

\subsubsection{FF Lens}

\begin{figure}
\includegraphics[scale=0.5]{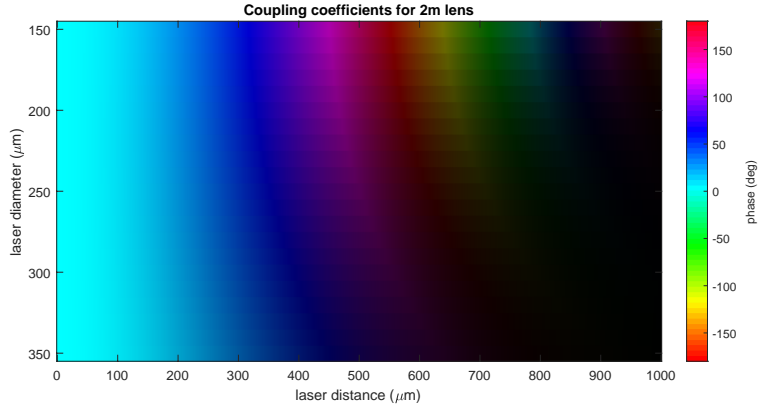}
\caption{\label{fig:ff lens coupling coeff}The coupling coefficients as a function of the distance between the lasers and their diameter, for a 2m FF lens. The color hue represents the phase of the coupling, and its brightness represents the normalized coupling strength.} 
\end{figure}

In the experiments where a FF lens was used as the diffractive element, we used the numerical simulation (figure \ref{fig:ff lens coupling coeff}) to find the lasers' diameter and distance that result in a negative coupling (180\textdegree) for a $2m$ lens. In the experiment we used the maximal laser diameter that allows single-mode lasing, and used figure \ref{fig:ff lens coupling coeff} to find the proper distance for negative coupling. We ignore next NN coupling due the Gaussian dependence of the coupling strength on the distance.

\subsubsection{Coupling matrix}
The coupling matrix of the laser network depends on the geometry of the network and diffraction pattern of the lasers after a single cavity RT. As explained above, the diffractive elements and the network geometry were chosen to satisfy negative coupling to NN, and negligible coupling to the other lasers. Hence, the coupling matrix of the laser network is given by the following matrix:
\begin{equation}\boldsymbol{\hat{\kappa}}=-|\kappa| 
\left(\begin{array}{ccccc}
0 & 1 & 0 & 1 & 1\\
1 & 0 & 1 & 0 & 1\\
0 & 1 & 0 & 1 & 0\\
1 & 0 & 1 & 0 & 0\\
1 & 1 & 0 & 0 & 0
\end{array}\right) \end{equation}
 where $|\kappa|$ depends on the details of the FF diffractive element and on the network geometry.

\subsection{Laser Intensity Control Scheme}   

To find the loss vector that yields uniform amplitude mode, we employed a closed-loop intensity feedback. The intensity feedback protocol is the following: Operating at quasi-CW mode, the pulse averaged intensity of each laser is measured using the imaging apparatus. This intensity is averaged over a pre-defined number of pulses
after which the SLM reflectivity of each laser is modified
according to a proportional control feedback scheme (eq. \ref{eq:intensity feedback})
with a constant target intensity for all lasers. If the reflectivity of one of the laser exceeds unity for five consecutive iterations, the target intensity is reduced by a fixed amount. This protocol is carried out until all intensities converge to the target value (up to a to a predefined tolerance).

\begin{figure}
\includegraphics[scale=0.4]{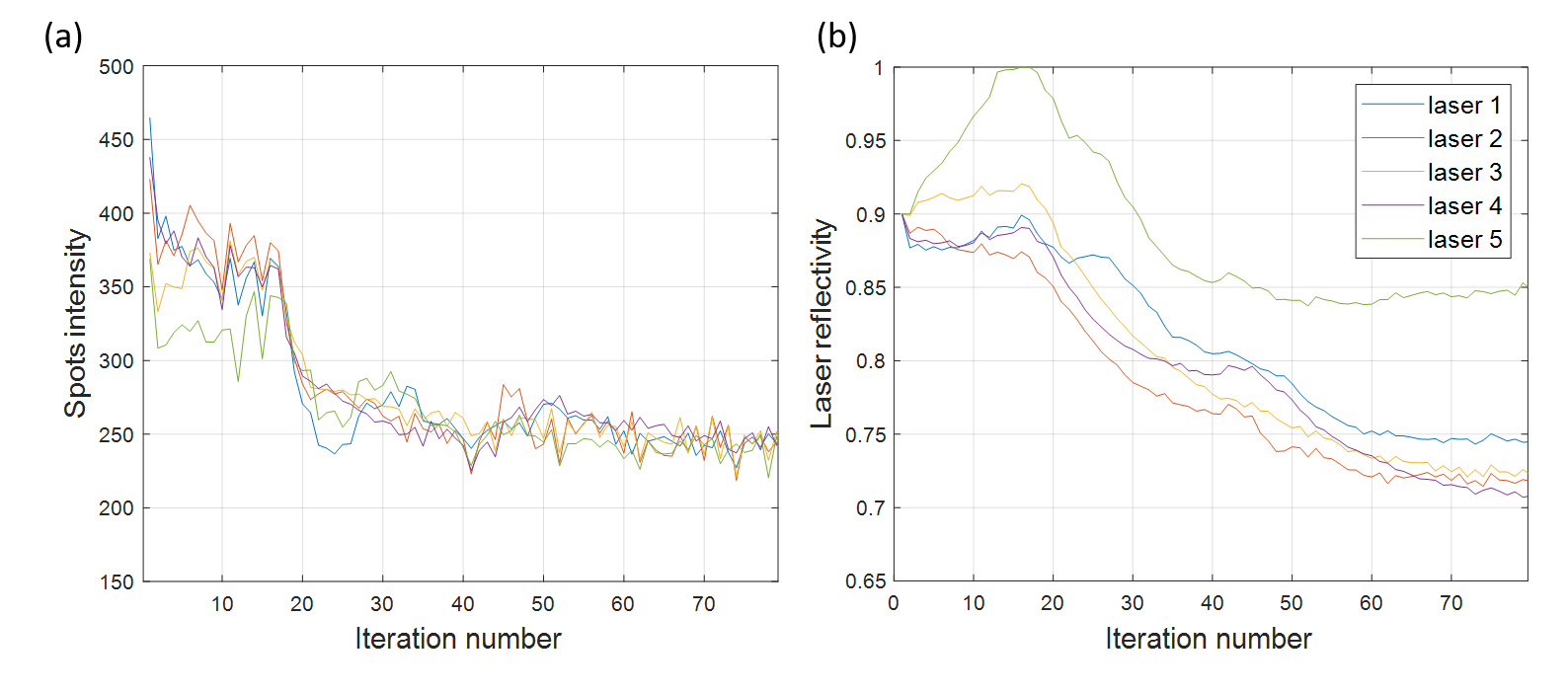}
\caption{\label{fig:intensity_feedback}Example of an intensity feedback trace on 5 lasers network (the House graph). Panel (a) displays the averaged intensity of each of the lasers as a function of the iterations. Panel (b) displays the SLM updated reflectivity for each of the lasers in each iteration} 
\end{figure}

In each feedback iteration, the reflectivity vector $\boldsymbol{r}_n$ is modified according to the following equation:
\begin{equation}\label{eq:intensity feedback}
\log\left(\boldsymbol{r}_{n+1}\right)=\log\left(\boldsymbol{r}_{n}\right)+K_{p}\cdot\left(\frac{I_{target}-\boldsymbol{I_{n}}}{I_{target}}\right)
\end{equation}
Where $n$ is the iteration number, $K_p$ is the proportional gain coefficient, $I_{target}$ is the intensity of the weakest laser in the first iteration, and $\boldsymbol{I}_{n}$ is the intensity vector. Figure \ref{fig:intensity_feedback} displays the intensity-feedback trace, for five laser on a House graph.

\subsection{Laser Frequency Control Scheme}

To achieve the required phase measurement resolution, the laser frequency heterogeneity i.e. detuning imperfection was also addressed. This imperfection is caused by phase aberrations and cavity misalignment which lead to systematic phase deviations in the laser network fixed points. To compensate for this imperfection, a frequency closed loop control scheme is implemented through the SLM phase. 
The frequency closed loop control scheme is based on (i) the multi-longitudinal-mode nature of the DDCL and (ii) the degeneracy of complex-conjugate solutions in a network of identical-frequency oscillators with symmetric real-values coupling matrix. These two properties result in real-valued coherence: a state with phases $\phi_{m}$ would give rise to a measured coherence of $\left\langle C_{m}\right\rangle = \frac{1}{2}\left(e^{i\phi_{m}}+e^{-i\phi_{m}}\right)=\cos\left(\phi_{m}\right)$.

Frequency detuning between the lasers changes the network fixed points such that the same phase is added to the ideally complex-conjugate solutions, and therefore leads to a complex-valued coherence. The detuning is then compensated by an iterative change of the phases on the SLM for the individual lasers, until real valued coherence is measured for all the lasers.

\subsection{Experimental Imperfection Compensation} 


Due to experimental imperfections, making the oscillators identical entails the nullification of parasitic loss and phase variations. To that end, we applied the intensity and frequency feedback schemes separately to two sub-graphs of the house graph: a square graph and a triangle graph. Since both subgraphs have uniform intensity fixed points when the oscillators are identical, finding the loss values that result in uniform intensity compensates for parasitic loss heterogeneity. The loss values resulting in uniform intensity are found by using the closed loop intensity feedback as described in previous section, and the phase values that result in real-valued coherence, are found by frequency feedback. In the first step of the compensation procedure, intensity feedback and frequency feedback were simultaneously applied to the square subgraph. In the second step, the two feedback loops were applied to the triangle graph, with fixed reflectivities and phases for the two base lasers  with the values from the first step (i.e. only the complex reflectivity of the apex laser was modified in the feedback). 


\subsection{House Graph XY model Ground State} 

In the XY ground state we have $ \phi_{12}=\phi_{23}=\phi_{34} \equiv \phi_{\sqcup} $, $\phi_{01}=\phi_{40} \equiv \phi_{\wedge}$ and $\phi_{14}  \equiv \phi_{-}$, which are related by

\begin{figure}
    \centering
    \includegraphics[scale=0.4]{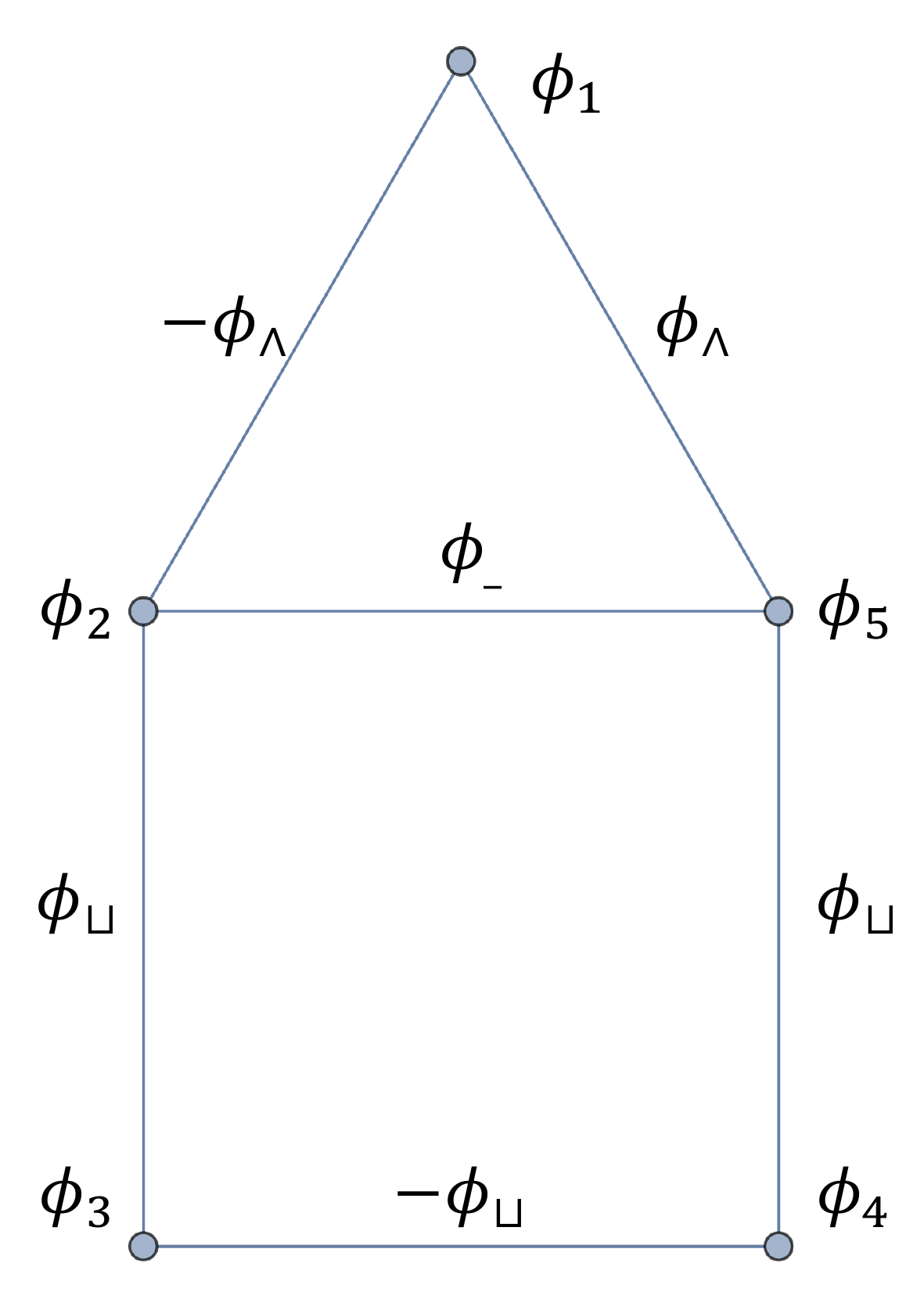}
    \caption{The house graph and notations for phase values and phase differences. Phase values are drawn next to vertices and phase differences are drawn next to edges}
    \label{fig:House_graph}
\end{figure}

\begin{eqnarray}
\sin \phi_{\sqcup} +\sin \phi_{-}   = \sin \phi_{\wedge}  \label{eh1}
\end{eqnarray}
and the geometric condition
\begin{eqnarray}
3 \phi_{\sqcup} - \phi_{-} =  2 \phi_{\wedge} + \phi_{-} = \pm 2 \pi \label{eh2}
\end{eqnarray}
where the $\pm$ is given by the sign of $\phi_{\wedge}$. 
Combining equations \ref{eh1} and \ref{eh2} leads to a single equation for $\phi_{\wedge}$, which is
solved by $ \phi_{\wedge} \approx \pm 0.62 \pi$.
The XY ground state solution is shown in figure 1 (b) in the manuscript.

\subsection{House Graph Exact Loss Vector} 

Here we demonstrate our approach for the house model. Using equation 8 in the main text, the added loss which corresponds to the XY ground state is uniquely defined by:
\begin{eqnarray}
 \Delta \alpha_0 \ &=& \ \delta - 2 \kappa \cos\phi_{\wedge} \\
 \Delta \alpha_1 \ &=&  \ \Delta \alpha_4 \ = \ \delta - \kappa \cos\phi_{\wedge} - \kappa\cos\phi_{\sqcup} - \kappa \cos\phi_{-}     \\
 \Delta \alpha_2 \ &=& \  \Delta \alpha_3 \ = \ \delta - 2 \kappa  \cos\phi_{\sqcup} 
\end{eqnarray}
A natural choice for $\delta$, is such that the minimal $\Delta \alpha_m$ is zero.
This way, for a given $P,\alpha$ and $\kappa$, we would achieve uniform intensity with the largest possible amplitude.
Here this corresponds to ${\Delta \alpha_0=0}$, so that $\delta= 2 \kappa \cos(\phi_{\wedge}) \approx -0.73 \kappa $.
Substituting $\delta$ and the phases we previously obtained leads to 
${\Delta \alpha_m \approx \kappa(0,1.3,1.2,1.2,1.3)}$.


\end{document}